\documentstyle[preprint,aps]{revtex}

\begin{document}
\draft

\title{\bf Theory of Spin-Split Cyclotron Resonance in the Extreme
Quantum Limit}
\author{N.R. Cooper and J.T. Chalker}
\address{Theoretical Physics, University of Oxford,\\
1 Keble Road, Oxford, OX1 3NP, United Kingdom.}

\date{\today}

\maketitle

\begin{abstract}
We present an interpretation of recent cyclotron resonance experiments
on the two-dimensional electron gas in GaAs/AlGaAs heterostructures.
We show that the observed dependence of the resonance spectrum on
Landau level occupancy and temperature arises from the interplay of
three factors: spin-splitting of the cyclotron frequency; thermal
population of the two spin states; and coupling of the resonances for
each spin orientation by Coulomb interactions. In addition, we derive
an $f$-sum rule which allows spin polarisation to be determined
directly from resonance spectra.
\end{abstract}


Since the discovery of the quantum Hall effect, there has been intense
interest in the properties of a two-dimensional electron gas in a
strong magnetic field.  Although it was the dramatic transport
phenomena that originally drew attention to these systems, increasing
use is being made of other experimental probes. At very low filling
fraction, transport measurements are problematic and optical
techniques such as photoluminescence and cyclotron resonance are
particularly valuable. The optical spectra show many interesting
features, and are still far from understood.

Recently, two groups have reported intriguing cyclotron resonance
measurements in GaAs inversion layers at low Landau level filling
fraction, $\nu$ \cite{exptold,exptnew}. For $\nu > 1/6$, a single
resonance peak is observed, which shows no anomalous behaviour at any
of the fractional quantum Hall states. Below $\nu \approx 1/6$, a
second resonance peak appears, and the absorption spectrum is strongly
dependent on both filling fraction and temperature.  These
observations were interpreted initially as a signature of Wigner
crystallization \cite{exptold}. Subsequently, it was demonstrated
\cite{exptold,exptnew} that the frequency splitting of the two
resonances at the lowest filling fractions is very similar to the
spin-splitting measured for cyclotron resonance in bulk GaAs, which
arises because the effective $g$-factor is energy-dependent
\cite{braun}. It has since been suggested that the observations may
indicate a spin-ordering transition, or localization of electrons in
the extreme quantum limit \cite{exptnew}.

In this paper we offer a detailed theoretical interpretation of these
experimental results. We show that the filling-factor dependence of
the cyclotron resonance spectrum is not due to any change in the
positional or magnetic correlations of the two-dimensional electron
gas. Instead, it is the result of a cross-over from independent,
spin-split resonances at low $\nu$, to a single mode dominated by
Coulomb interactions at high $\nu$. Moreover, we are able to account
for the temperature dependence of the spectrum in terms of the thermal
populations of the two spin states, calculated using the known
single-particle $g$-factor and without invoking exchange interactions.

Kohn's theorem \cite{kohn} provides a natural point of departure for
theories of cyclotron lineshape: in a translationally invariant,
one-component system, cyclotron resonance couples only to the centre
of mass, and the resonance is unaffected by interactions or
correlations.  Spectral structure may be observed if translational
invariance is broken, either by disorder \cite{kalperin} or by
non-parabolicity of the conduction band \cite{mackallin}.
Alternatively, it may arise if there are carriers of more than one
type, with different cyclotron frequencies, so that centre of mass
motion does not separate from internal motion of the electron
gas. This happens in strained Si inversion channels, in which separate
pockets of the conduction band are occupied, and has been studied
experimentally \cite{stallhofer} and theoretically \cite{appelover} at
weak magnetic field. There is a clear indication that the same
mechanism is at work in the experiments we are concerned with, from
the fact that the frequency splitting of the resonance at low $\nu$
matches the bulk spin-splitting \cite{exptnew}.

Motivated by this, we develop in the following a theory of cyclotron
resonance for a two-component system in the strong field limit, the
two components representing spins of each orientation. A complete
treatment of many-body correlations at finite temperature would
clearly be very difficult. Fortunately, the experimental conditions
suggest several simplifications.  First, the filling fractions of
interest are small, so exchange energies are negligible \cite{lam},
and the electrons may be treated as distinguishable particles, each
with a conserved spin orientation. Second, the cyclotron energy is
much larger than all other energy scales, so that only the two spin
states of the lowest Landau level are populated, and cyclotron
resonance involves transitions only into the corresponding spin states
of the next level. Since higher Landau levels do not contribute, the
two spin orientations may each be treated as if they have a parabolic
dispersion relation, with slightly different effective masses.  The
relative concentrations of these two types of particle are fixed
simply by their Boltzmann weights, in terms of the Zeeman splitting of
the lowest Landau level. The favoured spin state has the higher
cyclotron frequency, so the minority spins can be thought of as
randomly distributed `heavy' impurities in a sea of the majority spin
state.

At low density, the natural co-ordinates are the guiding centre and
orbital co-ordinates of the electrons
\begin{equation}
\mbox{\boldmath $R$}_i  = \mbox{\boldmath $r$}_i - \mbox{\boldmath $\rho$}_i 
 \; , \; a_i = 
\frac{\rho_i^x-i\rho_i^y }{\sqrt{2}l} \; , \; a_i^{\dag} = 
\frac{\rho_i^x+i\rho_i^y }{\sqrt{2}l} ,
\end{equation}
where $\mbox{\boldmath$\rho$}_i=(\mbox{\boldmath
$p$}_i+e\mbox{\boldmath $A$}_i)\times\hat{\mbox{\boldmath $z$}}/eB$,
and $l=\sqrt{\hbar/eB}$ is
the magnetic length. The only non-zero commutators are
\begin{equation}
[a_i,a_j^{\dag} ] = \delta_{ij} \; $ and $ \; [R_i^x,R_j^y] =
il^2\delta_{ij} .
\end{equation}
Cyclotron resonance measures the real part of the dynamical
conductivity, which is proportional to \cite{andofowlerstern}
\begin{equation}
\label{eq:correlation} S(\omega) = {1 \over \omega}  \int_{-\infty}^\infty
\sum_{i,j}\langle a_i(t)a_j^{\dag}(0) \rangle \mbox{e}^{i\omega t}
\;\mbox{d}t .  
\end{equation} 
In view of the low filling fraction, we expand the interactions in
powers of the orbital co-ordinates. At leading order in the
interactions, the equations of motion are
\begin{eqnarray} \label{eq:rdot}
\dot{\mbox{\boldmath $R$}_i} & = & -\frac{1}{eB} \hat{\mbox{\boldmath $z$}} 
\times \left.\frac{\partial U}{\partial \mbox{\boldmath $r$}_i}
\right|_{\{R_N\}}\\
\label{eq:adot}
\dot{a_i} & = & -i\omega_i a_i - \frac{i}{2eB} \sum_j a_j 
\left.\left(\frac{\partial^2 U}{\partial x_i \partial x_j} + 
\frac{\partial^2 U}{\partial y_i \partial y_j}\right)\right|_{\{R_N\}}
\end{eqnarray}
where $\omega_i$ is the cyclotron frequency of electron-$i$,
$U(\{r_N\})$ is the Coulomb energy, rapidly oscillating terms are
omitted and an additive constant has been absorbed in $a_i$.

Thus, from Eq~\ref{eq:rdot}, the guiding centres drift along contours
of the local potential gradient (independently of the orbital motion),
while, from Eq~\ref{eq:adot}, the orbital co-ordinates perform coupled
oscillations, with the coupling determined by the instantaneous
positions of the guiding centres. For Coulomb interactions the
coupling matrix is
\begin{equation}
\label{eq:matrix}
M_{ij} = \left\{ \begin{array}{ll} \sum_{k\neq i}
|\mbox{\boldmath $R$}_i- \mbox{\boldmath $R$}_k|^{-3} & i=j , \\ 
-|\mbox{\boldmath $R$}_i-\mbox{\boldmath $R$}_j|^{-3} & i\neq j .
\end{array} \right.
\end{equation}

At the lowest experimental temperatures and densities, the system is
expected to form a Wigner crystal, with the guiding centres located at
the sites of a triangular lattice.  In these circumstances, the
coupling matrix is time independent and the resonances are normal
modes satisfying
\begin{equation} \label{eq:nmode}
\omega a_i
= \delta\omega_i a_i + I \sum_j M_{ij} a_j ,
\end{equation}
where we measure frequency, $\omega$, relative to the cyclotron
frequency, $\omega_c$, of the majority spins, in units of the
splitting, $\delta\omega_c$, so that $\delta\omega_i = 0,-1$ for the
majority and minority spins, respectively. The dynamical matrix
(\ref{eq:matrix}) is defined by the triangular lattice vectors, in
units of the lattice constant, $a= (\sqrt{3}/2\:n)^{-1/2}$, and $I$ is
the dimensionless interaction strength
\begin{equation}
\label{eq:interaction} I =
\frac{e^2l^2}{8\pi\epsilon\epsilon_0 a^3} \times
\frac{1}{\hbar\delta\omega_c}.
\end{equation}

Hence we have established a model for cyclotron resonance in the
Wigner crystal which has just two parameters: the interaction
strength, $I$, controlled by the electron density; and the
concentration of the minority spin state, $p$, set by the ratio of
temperature to Zeeman splitting.  It is similar to models for the band
structure of uncorrelated binary alloys, and various approximate
methods of solution are available. To facilitate a direct comparison
with experiment, we choose, however, to solve Eq~\ref{eq:nmode} by
numerical diagonalization.

Before presenting the results of our calculations, it is appropriate
to review the experimental data, the most striking aspect of which is
the temperature dependence of the spectrum. At low density, two
resonance peaks are observed at the positions of the separate spin
transitions, Fig.~1(a). At a slightly higher density, the behaviour is
much more complicated: both peaks shift continuously with temperature
and eventually collapse together, Fig.~1(b). Above $\nu\approx 1/6$,
there is a single peak (not shown), whose position depends weakly on
temperature.

These three types of behaviour correspond to the weak, intermediate
and strong coupling regimes of our model. We have calculated the
absorption spectra in each regime for a system of 225 particles, using
periodic boundary conditions and Ewald summation for the dipole forces
\cite{lambin}, and averaging over 400 realizations of the spin
distribution. Longer studies of up to 900 particles show that finite
size and statistical errors are small under these conditions. We
display our results in Fig~2. At weak coupling, the two spins are
probed almost independently, Fig~2(a).  At intermediate coupling, the
resonance structure becomes strongly dependent on the spin
polarisation, controlled by temperature, Fig.~2(b).  In the strong
coupling regime (not shown), we obtain a single resonance at the
average cyclotron frequency, which shifts with temperature to follow
the changing spin population.

The successful reproduction in the calculated spectrum, Fig~2, of most
of the observed features, Fig~1, constitutes our main result. We
consider it persuasive evidence that our interpretation of the
experiments incorporates the relevant physical ingredients. In making
this comparison, the only scope for adjustment lies in the conversion
between experimental electron densities and the interaction strength,
$I$, of the model. The experimental densities of Fig~1 lead to values
for the interaction strength of $I=0.1, 0.17$, while we find the best
fit with smaller values: $I=0.05, 0.1$. We attribute the discrepancy
to the finite thickness of the electron gas, which softens the Coulomb
repulsion at short distances; comparable behaviour has been noted
previously in calculations on the fractional quantum Hall effect
\cite{fqhegap}. The spin populations are determined by the Zeeman
splitting, using the value \cite{gfactor}, $g=-0.4$, for the electron
$g$-factor,  and the theoretical spectrum is convolved with a
Lorentzian to match the experimental resolution.

It is initially surprising that the results of our calculations
represent so well the observations, since the theory assumes that
guiding centre coordinates lie on a lattice, while most of the
measurements are for temperatures well above the predicted melting
point of the Wigner crystal. The explanation, we believe, is that the
crossover from single-particle to collective cyclotron resonance is
relatively insensitive to interparticle correlations, because the
interactions are long-ranged. In principle, one could investigate
resonance in the liquid phase by using molecular dynamics to follow
guiding centre motion, simultaneously integrating Eqs~\ref{eq:rdot}
and \ref{eq:adot}. The data presently available does not justify such
an approach. Instead, we have examined the importance of interparticle
correlations by diagonalising Eq~\ref{eq:nmode} for {\it fixed} liquid
configurations of guiding centres, generated using a Monte-Carlo
method. This introduces a third parameter to the model,
$\Gamma=kT\,4\pi\epsilon\epsilon_0a/e^2$. In Fig.~3 we compare the
results, at temperatures corresponding to 4K and 20K for an electron
density of $3 \times 10^{10}$cm$^{-2}$, to those for the Wigner
crystal. The lineshape changes by less than the experimental
resolution, even at these high temperatures.

Further insight comes from studying a simplified model (suggested by
the long-range character of interactions) in which each electron is
coupled equally to all others, with a coupling constant inversely
proportional to the total number of particles.  This model is
obviously insensitive to the positions of guiding centres. It supports
only two optically-active modes, the in-phase and out-of-phase
oscillations of the two spin populations, with frequencies and
oscillator strengths given by
\begin{eqnarray}
\omega_{in} & = & (\alpha - 1 - \sqrt{(1-\alpha)^2+4\alpha p})/2 \\
\omega_{out} & = & (\alpha -1 + \sqrt{(1-\alpha)^2+4\alpha p})/2 \\
S_{in} & = & 1-S_{out} = \frac{\omega_{out}+p}{\omega_{out}-\omega_{in}}
\end{eqnarray}
where $p$ is the concentration of the minority spins. For comparison
with our previous results, we choose the coupling constant, $\alpha$,
so that the sum of the interaction matrix over all neighbours is the
same as that obtained with Coulomb interactions in the Wigner crystal,
which requires that $\alpha = 11.034I$. The frequencies and oscillator
strengths are displayed in Fig~4: as the coupling is varied, there is a
crossover from independent resonances of the two spin populations at
weak coupling, to a single, in-phase, mode which will lie at the average
cyclotron frequency, $-p$, at strong coupling.  The crossover is most
sharply defined for a single minority spin, $p=0$, when it marks the
transition from a bound state on the impurity spin (with relative
oscillator strength inversely proportional to total particle number),
to an extended state which carries all the oscillator strength. This
transition would not be present in a crystal with short range
interactions, since an arbitrarily weak  attractive potential always
has a bound state in two dimensions when the dispersion relation is
quadratic at small wavevector. It does, however, occur in the
two-dimensional Wigner crystal with Coulomb interactions, because the
magnetoplasmon dispersion relation is {\it linear} at small wavevector.
We find the critical interaction strength for the appearance of a bound
state to be $I_c=0.104$ in this case, while for infinite range
interactions $\alpha_c = 1$, implying $I_c = 0.091$.

Although spin-split cyclotron resonance is, as we have shown, very
insensitive to spatial correlations between electrons, it does provide
a direct measure of spin polarisation. We demonstrate this by
considering the first moment of the spectrum. To first order in
$\delta
\omega_c / \omega_c$
\begin{eqnarray}
\langle w\rangle & = & \left.\int_{-\infty}^\infty \omega S(\omega) \;
\mbox{d}\omega \right/
\int_{-\infty}^\infty S(\omega) \;\mbox{d}\omega \\
  & \simeq & \sum_{i,j} \left.\langle \omega_i a_i a_j^{\dag} + 
(l/\hbar)\partial_{i} U a_j^{\dag}\rangle \right/ \sum_{i,j} \langle 
a_i a_j^{\dag} \rangle
\end{eqnarray}
where $\partial_i = (\partial /\partial x_i +i\partial /\partial y_i)$
and $U$ is the full interaction. For pairwise interactions, $\sum_i
\partial_i U = 0$, and in the absence of
Landau level mixing, $\langle a_i a^{\dagger}_j \rangle = \delta_{i,j}$, 
so we find
\begin{equation} \label{eq:sum}
\langle w\rangle  = \frac{1}{N}\sum_{i=1}^N\omega_i.
\end{equation}
Thus the mean absorption frequency is simply the average cyclotron
frequency of the particles. Given the cyclotron frequencies for each
spin orientation, it provides a direct measure of the relative
populations.

In conclusion, we have presented a model for spin-split cyclotron
resonance at low filling fraction. Within this model we have been able
to account fully for recent cyclotron resonance measurements on GaAs
heterostructures. We find that the technique is rather insensitive to
the liquid-solid phase transition, but provides a direct measure of
spin polarisation.

We are grateful to R. J. Nicholas for many helpful discussions. This
work was supported in part by SERC grant Gr/Go 2727, and in part by
DENI.

\newpage

\newpage
\begin{center} FIGURE CAPTIONS \end{center}

FIG. 1. Temperature variation of the experimental resonance spectrum,
reproduced from [2]. The electron densities are (a) $2.6\times10^{10}
\mbox{cm}^{-2}$ and (b)  $3.4\times10^{10} \mbox{cm}^{-2}$. 

\vskip 5mm

FIG. 2. Predicted resonance spectrum of the Wigner crystal, (a) $I=0.05$, 
(b) $I=0.1$

\vskip 5mm

FIG. 3. Comparison of the resonance spectrum in the Wigner crystal
($\Gamma=0$) and in the liquid at two temperatures ($\Gamma=0.2, 1.0$), 
for $p=0.1$.

\vskip 5mm

FIG. 4. Crossover of the meanfield resonances, for $p=0,0.01, 0.05,
0.1, 0.2, 0.3,0.4,0.5$: (a) frequencies, (b) oscillator strengths.

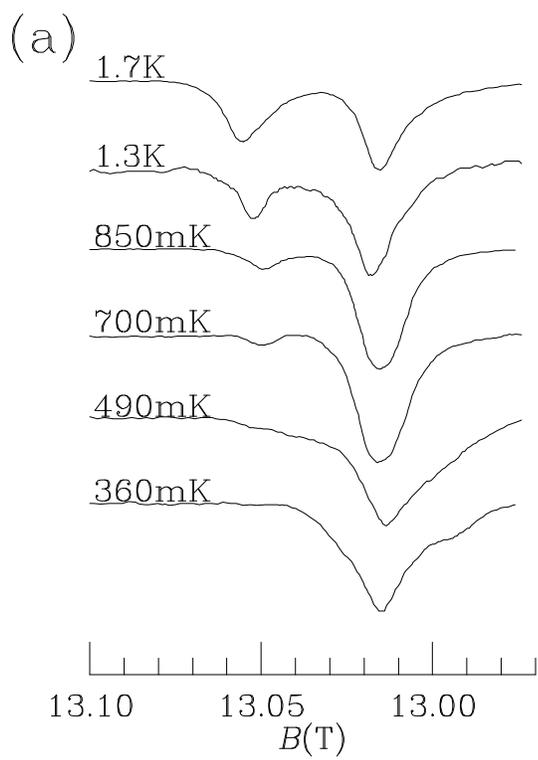
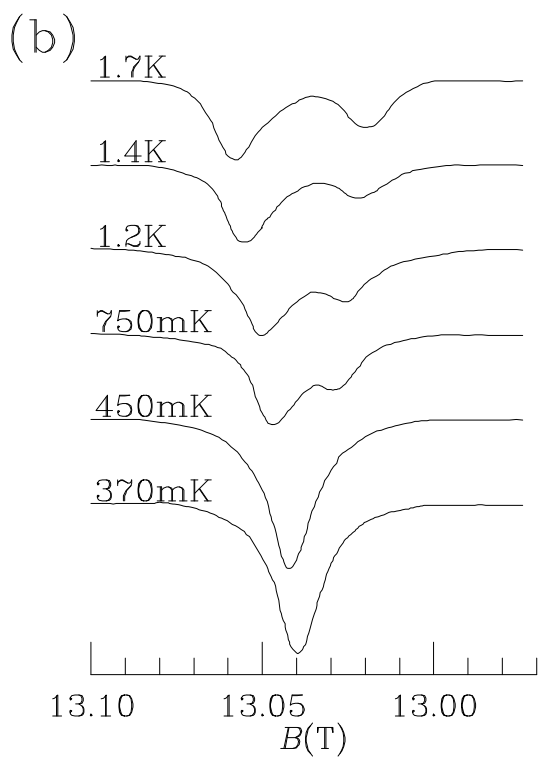
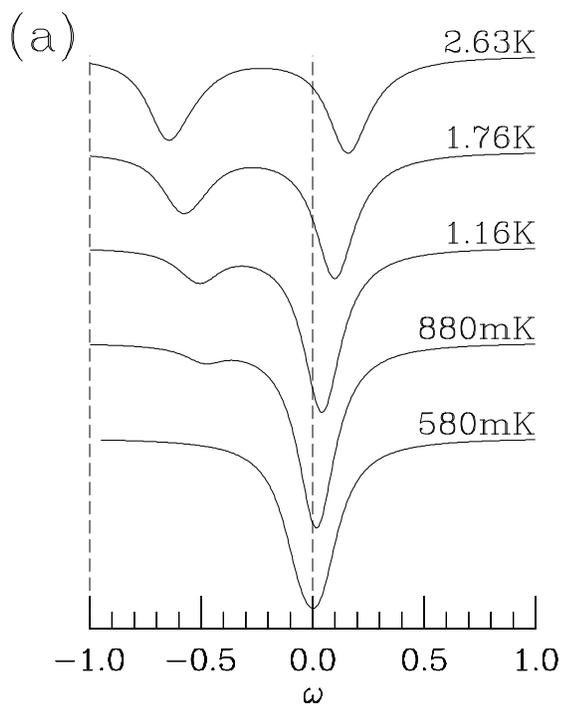
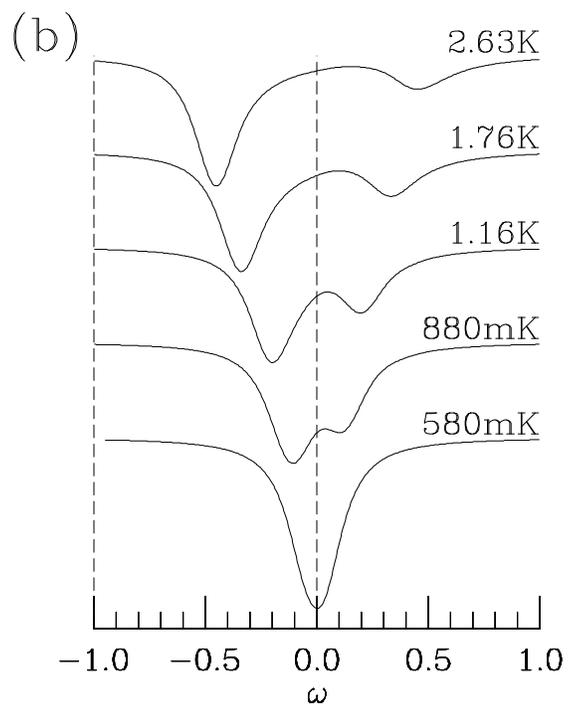
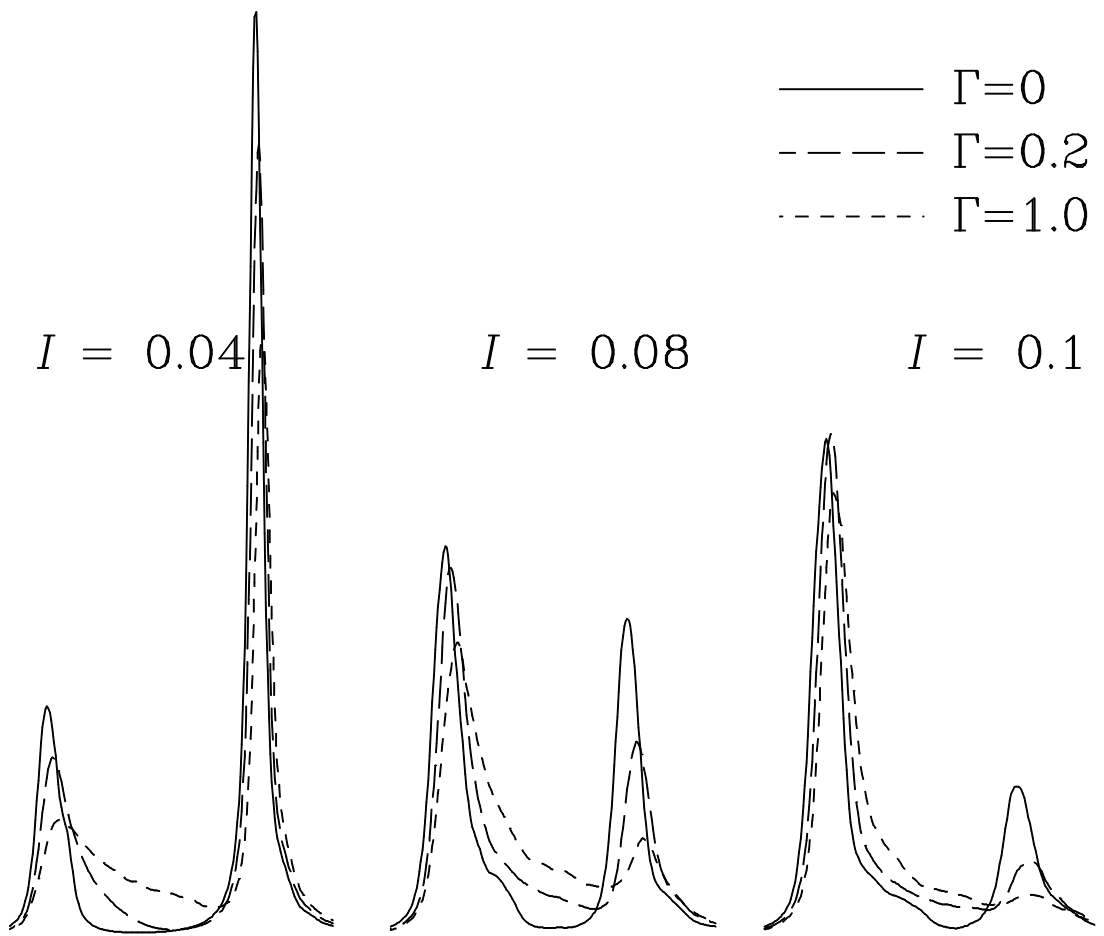
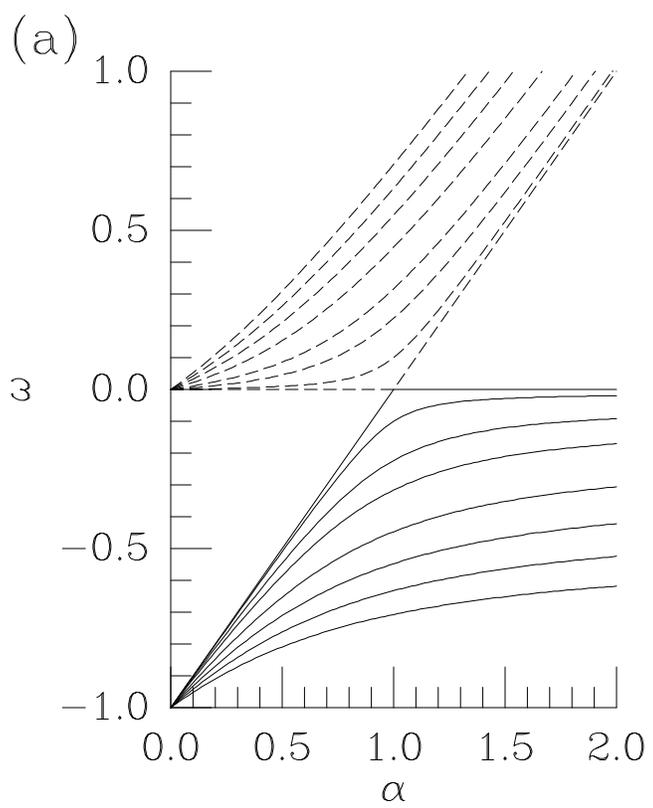
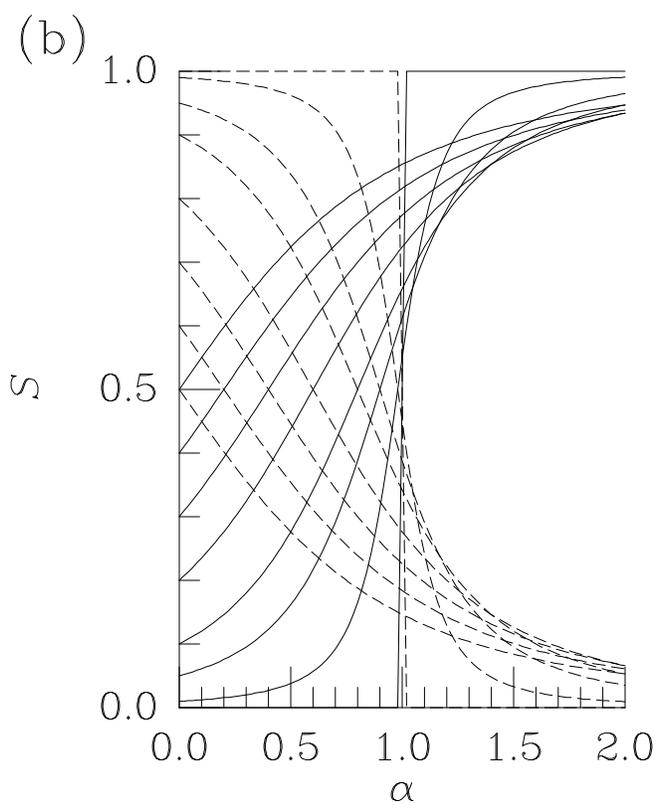
\end{document}